\documentstyle[prl,aps,epsf]{revtex}

\begin{document}
\draft

\title {
  Numerical Evidence for Spontaneously Broken\\
  Replica Symmetry  in 3D Spin Glasses}

\author{E. Marinari\cite{MAR}}
\address{Dipartimento di Fisica and Infn, Universit\`a di Cagliari\\
Via Ospedale 72, 09100 Cagliari (Italy)}

\author{G. Parisi\cite{PAR} and J. Ruiz-Lorenzo\cite{RUI}}
\address{Dipartimento di Fisica and Infn,
Universit\`a di Roma {\em La Sapienza}\\
P. A. Moro 2, 00185 Roma (Italy)}

\author{F. Ritort\cite{RIT}}
\address{Departamento de Matem\'aticas,
Universitad Carlos {\em III}\\
Butarque 15, Legan\'es 28911, Madrid (Spain)}

\date{\today}

\maketitle

\widetext

\begin{abstract}
By numerical simulations of the $3d$ Ising spin glass we find evidence
that spontaneous replica symmetry breaking theory and not the droplet
model describes with good accuracy the equilibrium behavior of the
system.
\end{abstract}


\narrowtext

The behavior of disordered spin models at equilibrium is well
understood in the framework of the mean field approximation
\cite{PARISI,PARBOO}.  The main prediction of the mean field approach
is the existence of a low temperature glassy phase. Such a phase is
characterized by the existence of many different equilibrium states
(spontaneous replica symmetry breaking, SRSB). On the other hand it is
possible to define a different consistent theory \cite{DROPLET} by
starting from the Migdal-Kadanoff approach.  By using a common
terminology we will refer in the following to this approach as to the
droplet model. Here one expects the equilibrium state to be unique
(apart from global inversions in zero magnetic field) and that the
most relevant excitations are obtained by reversing large domains of
spins (the droplets).

We have two different starting points. One is the infinite range
approximation which leads to the replica symmetry breaking picture and
the other is the Kadanoff-Migdal approximation which leads to the
droplet model. Although each of the two pictures is correct in its
range of validity we have to establish which of the two qualitatively
describes the physics of the real three dimensional spin glasses.

The main result of this work (which continues the investigation
started in \cite{MAPARI}, and follows a long series of Monte Carlo
simulations of spin glass systems \cite{MC}) has been to gather new
and strong evidence that in three dimensions the SRSB picture (and not
the droplet model) describes correctly what is observed in numerical
simulations.

Let us start by summarizing the evidence we will present in this note
and the scheme of our reasoning (for a more detailed exposition of
these and more data see ref. \cite{MPRR}). We will start by showing
that the probability distribution of the overlap among two systems at
equilibrium, $P(q)$, has a non-trivial structure.  $P(q\simeq 0)$ is
different from zero, and its shape does not depend on the volume size.
We will analyze (following a suggestion contained in the third
reference of \cite{CAR}) sample to sample fluctuations of the
spin-glass susceptibility, and find out that they are incompatible
with the droplet model, while their size is very well explained (even
in a quantitative manner) by SRSB theory.  In order to show that the
structure of the different equilibrium states is not compatible with a
droplet structure we will compute and analyze equal time correlation
functions. From this analysis we deduce the existence of many
equilibrium states that cannot be described by a droplet like
structure. On the contrary we will show that even at a quantitative
level SRSB theory explains very well the numerical data.

Further evidence about the inadequacy of the droplet model to describe
the $3d$ spin glasses and support to a SRSB mechanism will be provided
by analyzing the distribution of overlaps of boxes of side $R$, and by
discussing the behavior of the box overlap Binder parameter.

The model we will consider most is defined by the simple
Edwards-Anderson Hamiltonian on a $3d$ simple cubic lattice
$ H \equiv - \sum_{\{i,k\}} \sigma_i J_{i,k} \sigma_j$,
where the sum runs over nearest neighbor couples of sites.  The
quenched disordered couplings $J$ are distributed according to a
Gaussian law.  A study of the overlap susceptibility and of the Binder
cumulants shows that (under the a-priori assumption the existence of a
phase transition at a non-zero temperature with a power law
divergence) the transition is located at $T\approx 1$.  In order to
check universality of our results we have also studied a model
\cite{MPRR} with integer $J=\pm 1$ variables, where each spin is
coupled with equal strength to 26 neighboring sites (all the ones
contained in a cube of $3^3$ sites). The results we discuss here are
confirmed by our findings about this second model.

We have used an isotropic lattice of linear size $L$, and we have
computed the probability distribution $P_J(q)$ of the overlap $q\equiv
V^{-1}\sum_i \sigma_i \tau_i$ among two thermalized configurations
$\sigma$ and $\tau$ in a box of volume $V=L^3$.  We have studied the
behavior of the function $P(q)$ averaged over a large number of
realizations of the quenched disordered couplings $J$ (i.e. the
average over the $J$ random variables of $P_J(q)$). We have used a
maximum of $2560$ samples for the smallest lattice sizes and a minimum
of $512$ samples for the largest sizes.  It was already known (see for
example \cite{MAPARI} and references therein) that it is non-trivial
and it has a shape quite similar to the one predicted in the mean
field model.  Mainly thanks to the use of large computer resources (we
have mainly used the APE parallel computer \cite{APE}, which turns out
to be very effective for this kind of problems \cite{MPRR}: we flip
about $2 \cdot 10^8$ spins per second on the {\em tower} version of the
machine ) and of the {\em tempering} (an annealing-like improved Monte
Carlo technique introduced in \cite{TEMPER}) we have been able to
study systems of larger size than before (up to $14^3$), and to bring
them to thermal equilibrium quite deep in the cold phase.  In this
case we have equilibrated the system up to distance $14$.  We will see
that this information is complemented by our dynamical study, where we
work on time scales on which we can equilibrate the system on
distances up to order $6$. This gives a good control over the fractal
geometry of the typical excitations and of their boundaries. This is
what we need in order to distinguish between SRSB theory or
Migdal-Kadanoff droplets.

The first crucial comment is that the shape of the function $P(q)$ is
in our statistical precision size independent.  What we stress is that
all the non-zero {\em plateau} at low $q$ values, down to $q=0$, turns
out to be size-independent.  For example at $T=0.7$ the Binder
cumulant of $q$ is practically independent from the lattice size and
it is equal to $0.85\pm.01$.  This means that the system has a
non-trivial structure of equilibrium states with a continuous
distribution of the allowed overlap values. By using our measurements
of equal time correlation functions we will argue in the following
that such states cannot be described by the droplet approach, while
they have all features predicted by the SRSB approach.

In this note we do not answer a very important question,
i.e. if in the infinite volume limit a low-temperature phase
characterized by the existence of a non zero order parameter $q_{EA}$
exists. On the lattice volumes we are able to investigate the high $q$
peak of the $P(q)$ is very slowly shifting toward lower $q$ values,
even if, as we already said, the shape of the $P(q)$ does not
change. The extrapolation to the infinite volume limit looks in this
case a very delicate issue, and many potential systematic errors (even
in the definition of the finite volume $q_{EA}$) are involved.  Here
we will not address in detail this point, and assume that we are
working in conditions where the system is effectively frozen to a
phase with a non-zero value of $q_{EA}$. A possible scenario
\cite{MAPARI} of a correlation length diverging exponentially for
$T\to 0$ or of a Kosterlitz-Thouless like transition would be
compatible with this approach, since on our finite lattice we would be
measuring properties of a frozen system. It is also important to note
that this ambiguity only concerns the behavior of the high $q$ peak of
the $P(q)$ (which could tend to $q=0$ on very large lattices), while
on the contrary the $P(q)$ for small $q$ values is non-trivial and
does not depend on the lattice size.

The agreement with mean field theory becomes quantitative if we study
sample to sample fluctuations.  Mean field theory tells us how much the
function $P_J(q)$ for a given realization of the quenched disorder
differs from the average. For example if we consider
$ \langle q^k \rangle_J \equiv \int dq\ P_J(q)\  q^k$,
we have that in mean field

\begin{equation}
  \overline{ \langle q^k \rangle_J \langle q^m \rangle_J} \ =\
  \frac23 \overline{\langle q^k \rangle_J}\  \
          \overline{\langle q^m \rangle_J}
  \ +\ \frac13 \overline{\langle q^{k+m} \rangle_J}\ ,
  \protect\label{E_STS}
\end{equation}
where by the overline we indicate the average over the quenched noise.
We have verified that in the low $T$ region this equality is very well
satisfied. For example for $k=2$ and $m=2$
at $T=0.7$ and $L$ ranging from $4$ to $10$ the ratio of the l.h.s. to
the r.h.s. of (\ref{E_STS}) is equal to $1.0$ with an error never
larger than $0.1$.

Strictly speaking the non-triviality of the function $P(q)$ is not in
violent contradiction with the droplet model. In the framework of the
droplet approach it is always possible to suppose that states where
domains that take a finite part of the whole system are reversed have
a finite probability.  This hypothesis is however rather unnatural and
it is definitely wrong in the Kadanoff-Migdal approximation.  Moreover
we have already seen that the ability of the SRSB theory to predict
quantitatively the fluctuations of the function $P(q)$ is remarkable.

We will now falsify the possibility discussed in the last paragraph by
considering the $q-q$ correlation functions restricted to those pairs
of configuration which have a small value of $q$. The analysis of such
correlation functions, together with the non-triviality of the $P(q)$,
will constitute an ultimate test of the failure of the droplet model.

More precisely we consider a system of side $L$ and we define the
relevant correlation function as

\begin{equation}
C(x,L)=V^{-1} \overline{ \langle{\sum_i \sigma_{i+x}\tau_{i+x}\ \sigma_i
      \tau_i}\rangle}\ ,
\end{equation}
where the brackets indicate the thermal average.  The droplet model
predicts that $C(x,\infty)$ goes to the constant value $q_{EA}^2$ for
large $x$.  In the SRSB approach $C(x,\infty) \propto |x|^{-\lambda},$
where $\lambda$ in an appropriate exponent which has been computed in
less than $6$ dimensions for the $q=0$ correlation functions
\cite{DEDOMI}.

We have studied this problem by considering large systems, with
$L=64$. We have ran numerical simulations starting from two random
configurations selected independently (for $4$ realizations of the
quenched couplings).  We have verified that $q^2$ stays small in the
whole run so that the difference in the initial configurations, for
not too large times, enforces the condition $q\approx 0$. Eventually
in a finite system global equilibrium will be reached and $q$ will
become of order $1$. However if we let $L\to\infty$ first we can use
this approach to study the equilibrium value of the correlation
function with the constraint of having zero overlap.

\begin{figure}
\epsfxsize=250pt\epsffile{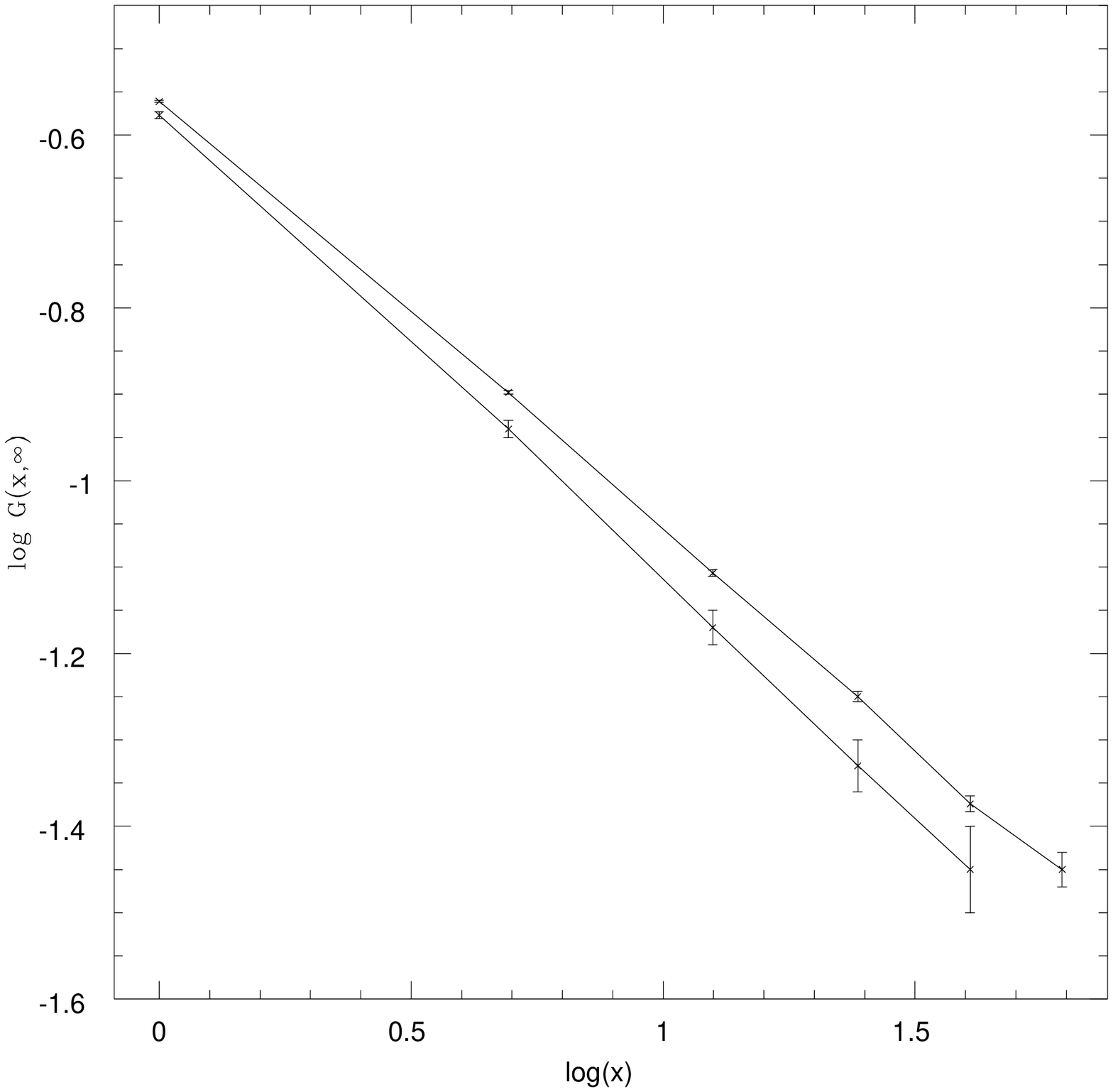}
\caption{
$\log{(G(x,\infty))}$ as a function of $\log{(x)}$.
The upper line from cooling, the lower one from the normal dynamics
(see the text for details).
}
\protect\label{F_TINF}
\end{figure}

In order to do that we consider the time dependent equal time
correlation function at time $t$

\begin{equation}
  G(x,t)= V^{-1}
  \overline{\sum_i
    \langle
      \sigma_{i+x}\tau_{i+x}\  \sigma_i \tau_i
    \rangle_t}\ ,
\end{equation}
where the average is done at time $t$, i.e.  after $t$ Monte Carlo
cycles after the random start.  We find that for large times $t$ the
correlation function $G(x,t)$ is essentially different from zero for
distances not too larger than a dynamic correlation lengthy $\xi(t)$
which increases (and maybe diverges) with time. Our numerical data are
well represented with the functional form

\begin{equation}
  G(x,t) = \frac{A(T)}{x^{\alpha}}\
  \exp \Bigl\{ - \bigl( \frac{x}{\xi(T,t)} \bigr)^\delta \Bigr\}\ ,
\end{equation}
where we have defined $ \xi(T,t) \equiv B(T)\ t^{\lambda(T)}$.  In the
whole range of distances $1\le x\le 8$ for Monte Carlo times which
range from $10^2$ to $10^6$ full lattice sweeps and a large range of
temperatures $T<T_c$ (we have done measurements at different
temperatures, down to $T_{min}\simeq .3 T_c$) we get good fits.  The
exponents $\alpha$ and $\delta$ are weakly dependent on $T$. For
example at $T=0.70$ we get the best values $\alpha = 0.50 \pm 0.02$
and $\delta=1.48 \pm 0.02$.  The correlation length exponent
$\lambda(T)$ is approximately given by $0.16 T$.  Such power law
growth of the correlation length was already observed by Rieger
\cite{RIEGER}.  In order to study the limit $t \to \infty$ in a safe
way it is even better to avoid global fits and to fit the data at
fixed distance $x$ as

\begin{equation}
  G(x,t)= G(x,\infty)\
  \exp \Bigl\{ -A(x) \ t^{-\lambda(T)}\Bigr \}\ .
  \protect\label{E_FIT}
\end{equation}
In this way the extrapolation to infinite time (with the
self-implemented constraint of $q=0$ always satisfied) is performed in
a very safe way. We plot in Fig. (\ref{F_TINF})
the correlations $G(x,\infty)$ (computed at $T=0.7$)
as a function of the distance in
double logarithmic scale.

We have also computed the same quantities by using a different
temperature schedule. In this second numerical experiment we slowly
cool down the system from $T=1.5>T_c$ to the final temperature. To
perform the cooling we use a number of steps proportional to $t$, the
waiting time we want to look the correlation function at. After that
the system evolves at the fixed temperature of interest $T$ for $t$
more time steps before measurement. In this way one can obtain a much
better equilibration. As matter of principle in this case one does not
expect a pure power law but a combination of different powers
generated by different temperature contributions.  However a fit
similar to the previous one (\ref{E_FIT}) works very well with a
slightly large value of of $\lambda$. At $T=0.7$ one obtains the
results shown in Fig. (\ref{F_TINF}). The data obtained with the two
techniques behave in a very similar way.  The $t=\infty$ data are well
described by a power decay $x^{-\alpha}$ with $\alpha=0.50\pm .03$, as
predicted by the replica theory and in variance with the droplet model
predictions.

The value of the correlation function at distance $1$ is particularly
interesting. Indeed in the model with Gaussian quenched disorder one
can easily prove by a simple integration by part that
$ E = - \beta (1 -C(1))$,
where $E$ is the energy per link and $C(x)$ is the overlap correlation
function of the fully equilibrated system (i.e. summed over different
ergodic components). Only if replica symmetry is broken it differs by
our correlation function $G$ where the two replicas have been kept (by
their own will) in two different ergodic components with zero mutual
overlap.  The energy can be computed with high accuracy and from its
value we can deduce $C(1)$. The equality

\begin{equation}
  C(1)=G(1)
  \protect\label{E_EQU}
\end{equation}
should be violated as soon as replica symmetry is broken, in the same
way in which the equality $E=-\frac{\beta}{2}(1-q_{EA}^2)$ is violated
in the Sherrington-Kirkpatrick model at low temperature.

The value of the energy per link is very well fitted by the form
$E_\infty + A t^{-\Delta(T)}$. The exponent $\Delta(T)$ turns out to
be quite large, i.e. we find   $\Delta(T)\simeq 0.44 T$, so that it is
not difficult to extrapolate the value of the energy to infinite time.
If we use the computation of the interface energy done by using SRSB
theory \cite{FRPAVI} we expect that $\Delta(T) = 2.5 \lambda(T)$,
which is very well satisfied by our data.

While we find that the equality (\ref{E_EQU}) is correct above and at
the critical temperature (with less that a relative $1\%$ error), it
is definitely violated below $T_c$; at $T=0.7$ we find $C(1) = 0.612
\pm .001$ and $G(1) = 0.56 \pm 0.01$, at $T=0.35$ we find $C(1) =
0.802 \pm 0.001$ and $G(1) = 0.67 \pm 0.01$.  The failure of the
equality (\ref{E_EQU}) implies the existence of different ergodic
components. The $q-q$ correlation function depends on the choice of
the component, in agreement with the main prediction of the SRSB
theory.

As a last evidence we discuss the value of the local overlap in a box
of side $R$, $ q_R(x) \equiv R^{-D} \sum_y \sigma_{x+y}\tau_{x+y}$,
where $y$ is an integer vector which takes all the $R^D$ values
compatible with the conditions $0 \le y_\nu < R$. We evaluate the
probability distribution $P_R(q_R)$ of the local box overlap.  In the
mean field SRSB limit the function $P_R(q_R)$ is Gaussian, but in a
finite (not too large) number of dimensions it is quite natural to
expect deviations from the Gaussian limit.

On the contrary in the droplet model the function $P_R(q_R)$ should have
two peaks at $q_R\approx q_{EA}$, and should become the sum of two delta
functions in the limit $R\to \infty$ (at least for $R\ll L$). Indeed
here the quantity $q_R$ is different from $q_{EA}$ with a probability
that goes to zero as a power of $R$.

\begin{figure}
\epsfxsize=250pt\epsffile{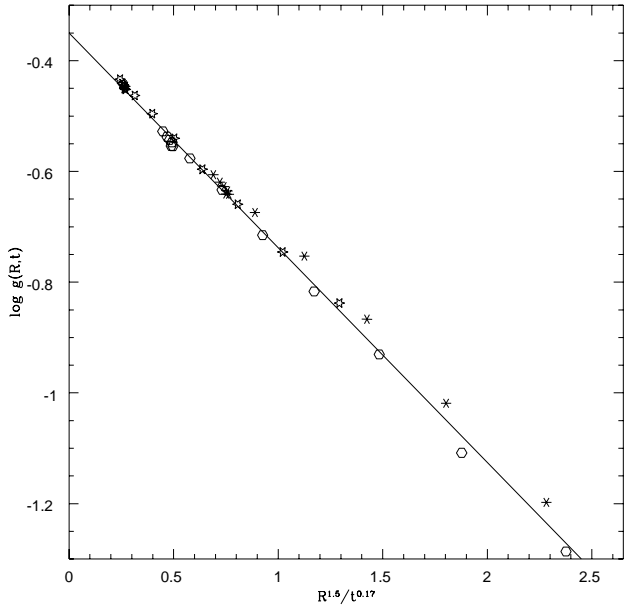}
\caption{
The logarithm of the Binder parameter for the box overlap versus
rescaled ratio of time and distance. Stars are for $R=2$, hexagons are
for $R=3$ and asterisks for $R=4$. The straight line is only to guide
the eye.
}
\protect\label{F_BIN}
\end{figure}

We have focused on the Binder parameter $ g(R,t) \equiv \frac32
-\frac{ \langle q_R^4\rangle} {2 \langle q_R^2 \rangle^2} $ measured
after $t$ Monte Carlo steps.  At a given temperature we expect that
for large $R$ the data will collapse as

\begin{equation}
  g(R,t) = f\Bigl( \bigl( R\xi(t)^{-1} \bigr)^{\delta}\Bigr)\ .
\end{equation}
In Fig. (\ref{F_BIN}) we show data for $T=0.7$
from $R=3$ and $4$.
The scaling law we are proposing works very well.  The
Binder function extrapolate to something definitely different from
$1$, which would be the prediction of the droplet model, since in that
case the distribution should be asymptotically a pair of delta
function. The data obtained with the alternative temperature
scheduling, by relatively slow cooling, give similar results for the
extrapolated value of the Binder cumulant.

We can summarize by saying that none of our findings give support to
the predictions of the droplet model, while the broken replica
approach is able to predict qualitatively (and in a few cases even
quantitatives) the behavior of the $3d$ spin glass systems.

We thank J. P. Bouchaud for pointing out to us the potential relevance
of local overlaps, and the {\em Ape} group for continuous assistance
and support.

\end{document}